# Semi-analytic Evaluation of 1, 2 and 3-Electron Coulomb Integrals with Gaussian expansion of Distance Operators W= $R_{C1}^{-n}R_{D1}^{-m}$, $R_{C1}^{-n}r_{12}^{-m}$, $r_{12}^{-n}r_{13}^{-m}$

Sandor Kristyan

*Research Centre for Natural Sciences, Hungarian Academy of Sciences,
Institute of Materials and Environmental Chemistry
Magyar tudósok körútja 2, Budapest H-1117, Hungary,*

Corresponding author: kristyan.sandor@ttk.mta.hu

**Abstract**. The equations derived help to evaluate semi-analytically (mostly for k=1,2 or 3) the important Coulomb integrals $\int \rho(\mathbf{r}_1)\ldots\rho(\mathbf{r}_k) W(\mathbf{r}_1,\ldots,\mathbf{r}_k) d\mathbf{r}_1\ldots d\mathbf{r}_k$, where the one-electron density, $\rho(\mathbf{r}_1)$, is a linear combination (LC) of Gaussian functions of position vector variable $\mathbf{r}_1$. It is capable to describe the electron clouds in molecules, solids or any media/ensemble of materials, weight W is the distance operator indicated in the title. R stands for nucleus-electron and r for electron-electron distances. The n=m=0 case is trivial, the (n,m)=(1,0) and (0,1) cases, for which analytical expressions are well known, are widely used in the practice of computation chemistry (CC) or physics, and analytical expressions are also known for the cases n,m=0,1,2. The rest of the cases – mainly with any real (integer, non-integer, positive or negative) n and m - needs evaluation. We base this on the Gaussian expansion of $|r|^{-u}$, of which only the u=1 is the physical Coulomb potential, but the u≠1 cases are useful for (certain series based) correction for (the different) approximate solutions of Schrödinger equation, for example, in its wave-function corrections or correlation calculations. Solving the related linear equation system (LES), the expansion
$$|r|^{-u} \approx \sum_{k=0}^{L} \sum_{i=1}^{M} C_{ik} r^{2k} \exp(-A_{ik} r^2)$$
is analyzed for $|r| = r_{12}$ or $R_{C1}$ with least square fit (LSF) and modified Taylor expansion. These evaluated analytic expressions for Coulomb integrals (up to Gaussian function integrand and the Gaussian expansion of $|r|^{-u}$) are useful for the manipulation with higher moments of inter-electronic distances via W, even for approximating Hamiltonian.

**Keywords.** Semi-analytic evaluation of Coulomb integrals for one, two and three-electron operators,
Higher moment Coulomb operators $R_{C1}^{-n}R_{D1}^{-m}$, $R_{C1}^{-n}r_{12}^{-m}$ and $r_{12}^{-n}r_{13}^{-m}$, with any real n, m,
Comments on programming real space incomplete gamma functions and reviewing vital properties of Gaussians

## INTRODUCTION

We evaluate the general Coulomb integral (most importantly for k=1,2 or 3)
$$\int \rho(\mathbf{r}_1)\ldots\rho(\mathbf{r}_k) W(\mathbf{r}_1,\ldots,\mathbf{r}_k) d\mathbf{r}_1\ldots d\mathbf{r}_k . \qquad (1)$$
For electron-electron (or nuclear-electron) interactions, the exact theory says that the Coulomb interaction energy is represented by the two-electron (or nuclear-electron) energy operator $W(\mathbf{r}_1,\mathbf{r}_2)\equiv W(1,2)= r_{12}^{-1}$ (or $W(\mathbf{r}_1)\equiv W(1)= R_{A1}^{-1}$) with the true physical $\rho(\mathbf{r}_1)$, while other W distance operators (mathematically weight functions) are useful to correct the approximate solutions of the Schrödinger equation with certain algorithms. In practice, LC of Gaussian type atomic orbitals (GTO) functions are used for approximation such as $\rho(\mathbf{r}_1)\approx \Sigma_A c_A G_{A1}$, in which
$$G_{Ai}(a,nx,ny,nz)\equiv (x_i-R_{Ax})^{nx} (y_i-R_{Ay})^{ny} (z_i-R_{Az})^{nz} \exp(-a|\mathbf{r}_i-\mathbf{R}_A|^2) \qquad (2)$$
with a>0 and nx, ny, nz ≥0 benefiting its important property such as $G_{Ai}(a,nx,ny,nz)G_{Bi}(b,mx,my,mz)$ is also (a sum of) GTO. (We use double letters for polarization powers i.e., nx, ny and nz to avoid "index in index", nx=0,1,2,… are the s, p, d-like orbitals, resp., etc.. In CC the nx, ny and nz goes up to 3 or 4.) The Coulomb interaction energy for molecular systems is expressed finally [1] with the LC of the famous integrals $\int G_{A1}G_{B2} r_{12}^{-1}d\mathbf{r}_1 d\mathbf{r}_2$ and $\int G_{A1}R_{C1}^{-1}d\mathbf{r}_1$. For corrections (correlation calculations, etc.) integrals such as $\int G_{A1}R_{C1}^{-2}d\mathbf{r}_1$, $\int G_{A1}G_{B2} r_{12}^{-2}d\mathbf{r}_1 d\mathbf{r}_2$, $\int G_{A1}G_{B2}G_{C3}r_{12}^{-n} r_{13}^{-m}d\mathbf{r}_1 d\mathbf{r}_2 d\mathbf{r}_3$, or with the more general W in the main title, etc. are also important. The value of n and m can be small positive integers, like W=1/($r_{12}r_{13}$) in the approximation by applying the Hamiltonian (**H**) twice for an optimized single determinant ground state wave-function ($S_0$) [2] ($E_{0,electr}\approx <S_0|\mathbf{H}^2|S_0>^{1/2}$), and negative, for example, in the so called R12 theory (see equation 52 in ref.[3]) the W= $r_{12}/r_{13}$, more generally, n and m can be any real number as a tool for correction.

Furthermore, GTO basis sets can usefully be augmented with correlation factors $\exp(-\gamma r_{ij}^m)$ with m=2 to improve the description of electron correlation effects; m=1 or 2 are the Slater or Gaussian-type geminals [4], resp.. Recall the ground state anti-symmetric analytic solution (with spin functions) of two-electron nucleus-free Shrödinger equation, $[(-1/2)(\nabla_1^2+\nabla_2^2) + r_{12}^{-1}]\Psi_k = E_{k,electr}\Psi_k$, which is $\Psi_0 = (\alpha_1\beta_2 - \alpha_2\beta_1)\exp(r_{12}/2)$ with $E_{0,electr}$=-0.25 [5].

Below, we use common notations, abbreviations and definitions: $F_L(v) \equiv \int_{(0,1)} \exp(-vt^2) t^{2L} dt$, the Boys function, L=0,1,2,…, $F_0(v) = (\pi/(4v))^{1/2} \text{erf}(v^{1/2})$; GTO= primitive Gaussian-type atomic orbital, the $G_{Ai}(a,nx,ny,nz)$ in Eq.2; $\mathbf{R}_A \equiv (R_{Ax}, R_{Ay}, R_{Az})$= 3 dimension position (spatial) vector of (fixed) nucleus A; $R_{AB} \equiv |\mathbf{R}_A - \mathbf{R}_B|$= nucleus-nucleus distance; $R_{Ai} \equiv |\mathbf{R}_A - \mathbf{r}_i|$= nucleus-electron distance; $\mathbf{r}_i \equiv (x_i, y_i, z_i)$= 3 dimension position (spatial) vector of (moving) electron i; $r_{ij} \equiv |\mathbf{r}_i - \mathbf{r}_j|$= electron-electron distance. The $G_{Ai}$ is called primitive Gaussian in CC, its simplest case is the nx=ny=nz=0 (to approximate 1s-orbitals ($\exp(-Z|\mathbf{r}_i - \mathbf{R}_A|)$)), denoted as $g_{Ai}(a) = G_{Ai}(a,0,0,0)$.

The analytic evaluation of (n,m)= (0,1) or (1,0) in the main title was fundamental and a milestone in CC and has a vast literature, we mention only one textbook [1] and two reviews [4,6]. These devices cannot be extended easily and systematically to general (n,m) values, except for n,m=0,1,2, see ref.[2]. The method described here is not only a procedure for any real (n,m), but also an alternative solution for the known n,m=0,1,2 cases, which serve as test. The simplest cases of Eqs.1-2 include the approximate 1s-orbitals (the simple Gaussian functions $g_{Ai}(a) = \exp(-a|\mathbf{r}_i - \mathbf{R}_A|^2)$):

$$\int_{(R3)} g_{P1}(p) \quad R_{C1}^{-1} \quad d\mathbf{r}_1 = (2\pi/p) F_0(p R_{CP}^2) \quad (3)$$

$$\int_{(R3)} g_{P1}(p) \quad R_{C1}^{-2} \quad d\mathbf{r}_1 = (2\pi^{3/2}/p^{1/2}) \exp(-p R_{CP}^2) F_0(-p R_{CP}^2) \quad (4)$$

$$\int_{(R6)} g_{P1}(p) g_{Q2}(q) \quad r_{12}^{-1} \quad d\mathbf{r}_1 d\mathbf{r}_2 = (2\pi^{5/2}/(pq)) \int_{(0,c)} \exp(-pq R_{PQ}^2 w^2) dw \quad (5)$$

$$\int_{(R6)} g_{P1}(p) g_{Q2}(q) \quad r_{12}^{-2} \quad d\mathbf{r}_1 d\mathbf{r}_2 = (2\pi^3 (pq)^{-1/2} (p+q)^{-1}) e^{-v} F_0(-pq R_{PQ}^2/(p+q)) \quad (6)$$

$$\int_{(R9)} g_{P1}(p) g_{Q2}(q) g_{S3}(s) r_{12}^{-1} r_{13}^{-1} d\mathbf{r}_1 d\mathbf{r}_2 d\mathbf{r}_3 = (4\pi^{7/2}/(qs)) \int_{(0,1)} \int_{(0,1)} g^{-3/2} \exp(-f/g) du\, dt \quad (7)$$

with $c \equiv (p+q)^{-1/2}$ in Eq.5 [4] and with $f \equiv pq R_{PQ}^2 u^2 + ps R_{PS}^2 t^2 + qs R_{QS}^2 u^2 t^2$ and $g \equiv p + qu^2 + st^2$ in Eq.7 [2]. Notice that product separation in $W = W_a(1) W_b(2) W_c(3)$ breaks Eq.1 into product of $\int \rho(\mathbf{r}_1) W_{a\,or\,b\,or\,c}(1) d\mathbf{r}_1$, as well as if e.g. $W = r_{12}^2$ or $r_{12}^4$, etc., then Eq.1 also reduces to products of $\int(...) d\mathbf{r}_1$. We do not discuss Hermite Gaussians ($H_{Ai}(a,t,u,v) \equiv (\partial/\partial R_{Ax})^t (\partial/\partial R_{Ay})^u (\partial/\partial R_{Az})^v \exp(-a R_{Ai}^2)$) [4,6] here, which generate many primitive Gaussians (e.g. $H_{Ai}(a,2,0,0) = -2a G_{Ai}(a,0,0,0) + 4a^2 G_{Ai}(a,2,0,0)$, etc.) with good orthogonal and overlap properties to speed up the calculation by reducing the terms; see a summary and comparison on speed and number of operations in ref.[7]. We call the attention that a good approximation with parameter set for u=1 in Eq.8 below for the Hamiltonian operator $\mathbf{H} \equiv \Sigma_{i=1}^N (-\nabla_i^2/2 - \Sigma_{A=1}^M Z_A R_{Ai}^{-1} + \Sigma_{j=i+1}^N r_{ij}^{-1})$ with r:= $R_{Ai}$ and $r_{ij}$ could substitute the classical way of evaluation [6] of Coulomb integrals (Eqs.3 and 5), more, all integrals in Hartree-Fock Self consistent field formalism [1] with STO-3G, 6-31G**, etc. basis sets, needing only the simple elementary integral $\int_{(-\infty,\infty)} x^n \exp(-ax^2) dx$ for n≥0 and no erf(x) and special tricks.

## Gaussian expansion of the Coulomb interactions (distance operators)

The idea of general Gaussian expansion of Coulomb interaction $1/|r|$ comes as early as in ref. [8] for calculations in nuclear physics for range I=(0.000005A, 0.0003A) and L=0 in

$$|r|^{-u} \approx \Sigma_{k=0}^L \Sigma_{i=1}^M C_{ik} r^{2k} \exp(-A_{ik} r^2) \quad \text{in the range I= } (0 < b_1 \leq r \leq b_2) \text{ and } A_{ik}, u > 0. \quad (8)$$

Our experience is that for high quality fit for a broad range in Eq.8 needs L=1, but L>1 may not necessary for electronic structures in CC, see Eq.9 below, although for test we analyze L>1 cases also, see Eq.10 below. For simplicity, we will switch double indices to single ones as $C_{ik} r^{2k} \exp(-A_{ik} r^2) \to c_i r^{2k} \exp(-a_i r^2)$ and (L+1)M→M. In Eq.8 the unit is bohr (a.u., ≈0.5A), and e.g. 0< $b_1$= 0.01 <<1 is a very close to nucleus distance, while e.g. 1<< $b_2$= 25 is a "relatively far" distance for (electronic) Coulomb forces (keeping in mind that it owns the property called "infinite, never disappearing" force); lucky situation in CC is that 1/r comes up as $\exp(-a R_{Ai}^2)/r_{ij}$, etc., so a larger $b_2$ distance is also a damped region. The algebraically lucky situation in Eq.8 is that the parameter set $\{c_i\}$ for 1-dimension variable (r) function can be directly transferred to 3-dimension variable ($\mathbf{R}_{A1}$ or $r_{12}$ as r:= $|\mathbf{R}_{A1}|$ or $r_{12}$) approximating functions for $R_{A1}^{-u}$ and $r_{12}^{-u}$ by the spherical symmetry. Notice that set $\{c_i\}$ can be the same for $R_{A1}^{-u}$ and $r_{12}^{-u}$, the set only depends on u and the range of fit in Eq.8. (Notice the tiny difference that $r_1 = |\mathbf{r}_1| > 0$ is a radius while |r|>0 is the abs. values of a 1-dim. variable). However, one must keep in mind the global differences between the two sides in Eq.8, that is, on interval [0,1] and [1,∞) the integral of $1/|r|$ is ∞, while for $r^{2k} \exp(-ar^2)$ the integral on [0,∞) is always finite. Also, $\lim |r|^{-u} = \infty$ vs. $\lim \exp(-a_i r^2) = 1$ as r→+0. Notice that integrals like in Eqs.3-7 are finite in spite of that $R_{A1}^{-n}$ and $r_{12}^{-n}$ have infinite values at zero distances if n>0. In our procedure, the evaluation of Boys function which needs the erf(x) function (one term, Appendix 1) in e.g. Eqs.3-6 is substituted by M terms in the approximate expansion in Eq.8. In program languages (e.g. FORTRAN) the calculation of erf(x) is not only very fast, but very accurate, so e.g. LSF for Eq.8 has to be at least a good quality. In an LSF for Eq.8 the erf(x) function is called $\approx M^2$ times (see next), and the quality depends on the value of M and set $\{a_i\}$, however in fact, a good $(M,\{c_i\},\{a_i\})$ for Eq.8 is enough to be calculated once and forever in CC.

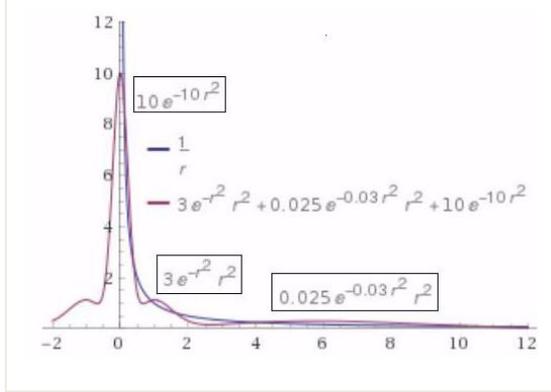

**FIGURE 1.** Displaying how the Gaussians in Eqs.8-10 recover $1/r^u$ in an interval I=(0<b1,b2). The dominant terms of LC in local regions (maximums) of I are shown in rectangular frames. The irrelevant (unphysical) negative (r<0) region is also shown where the LC completely leaves the Coulomb potential $1/r^u$. Notice the magnitude difference in $a_i$ values before and after r=1, i.e. 10 vs. 1 and 0.03, as well as how the decreasing $a_i$ =1 and 0.03 in $r^2\exp(-a_i r^2)$ shifts the location of maximum toward larger r. LC of $\exp(-a_i r^2)$ is dominant in $[b_1,1]$, but the $r^2\exp(-a_i r^2)$ in $[1,b_2]$.

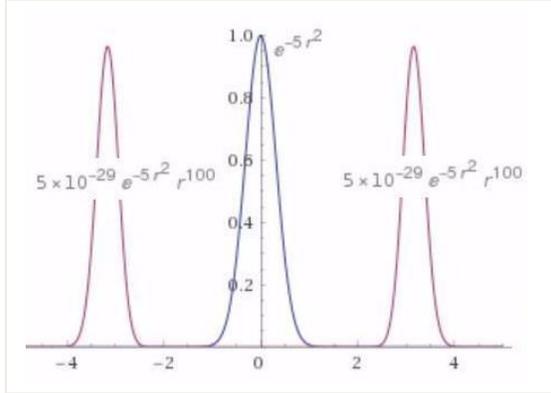

**FIGURE 2.** Displaying that the width$\approx (1/a_i)^{1/2}$ of terms $c_i\, r^{2k}\exp(-a_i r^2)$ in Eqs.8-10 depending very weakly on k, (but resulting extreme coefficients $c_i$ if k is large instead, as well as $r_{max}=\pm(k/a_i)^{1/2}$, see also Fig. 1). As a consequence, terms $\exp(-a_i r^2)$ and $r^2\exp(-a_i r^2)$ (k=0 and 1) may enough to approach $1/r^u$ adequately, and terms with large k values can be avoided in the approximation. Irrelevant region r<0 is also shown for the twin peak if k>0. (Miscellaneous is that y axis is the only symmetry element for k>0 too.)

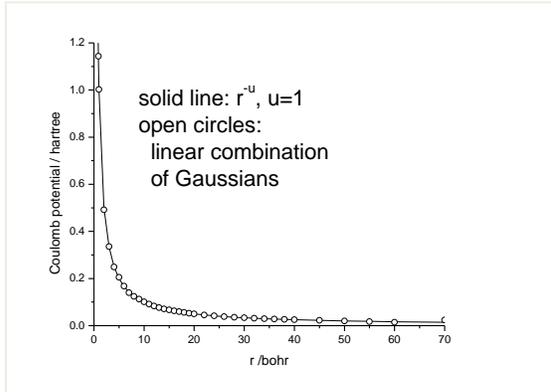 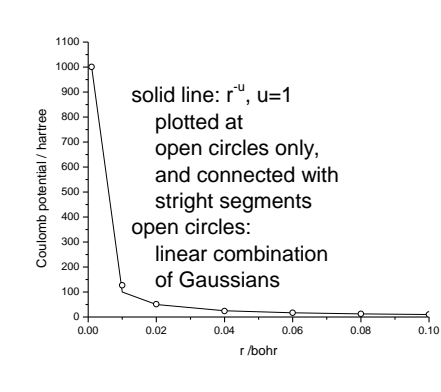

**FIGURE 3.** LSF (App. 2) of LC of Gaussians in Eq.9 to Coulomb potential 1/r displayed for ranges (0,70) and a scaled-up (0,0.1); 1/r alters from very steep to very flat at r=1.

The 2k power in Eq.8 ensures that substitution $r=(x_1^2+y_1^2+z_1^2)^{1/2}$ eliminates square roots, <u>fundamental in the next chapter!</u> There are two important subcases (using single index): the L=1 case

$$|r|^{-u} \approx f(r) \equiv \Sigma_{i=1}^{M0} c_i \exp(-a_i r^2) + \Sigma_{i=M0+1}^{M0+M2} c_i\, r^2 \exp(-a_i r^2) \quad \text{and } M=M0+M2 \quad (9)$$

as well as the L>>1 case, (using same meaning in power but different in sum for k in Eq.8 and 10 as)

$$|r|^{-u} \approx f(r) \equiv \Sigma_{i=1}^{M0} c_i \exp(-a_i r^2) + (\Sigma_{k=0}^{L} d_k\, (r^2-r_0^2)^k)(\exp(-a\, r^2)+ B \exp(-b\, r^2)) \quad \text{and } M=M0+L+1 \quad (10)$$

in which $r_0=1$ is chosen ($\Rightarrow |r_0|^{-u}=r_0^{2k}=1$). The $|r|^{-u}$ rapidly decreases in (0,1) but very flat in [1,∞), a property which must be considered seriously in the fit for Eq.8. The $M_0$ in Eqs.9-10 cannot be zero, at least M0=1 must be, since the k>0 terms in Eqs.9-10 are zero at r=0 in contrast to $|r=0|^{-u}=\infty$. The maximum of $\exp(-a_i r^2)$ is at r=0, so its inflection point at $r_{inf,i}=\pm(2a_i)^{-1/2}$ is chosen for set $\{a_i\}$ for i=1,… M0 by the uniform division of $I_1=[b_1,1)$, that is $r_{inf,i}:= b_1+(i-1)d$ with $d=(1-b_1)/M0 \Rightarrow a_i:= 1/(2r_{inf,i}^2)$ for i=1,2,…M0. The maximum of $r^2\exp(-a_i r^2)$ is at $r_{max,i}=\pm(a_i)^{-1/2}$, and similarly, $r_{max,i}:= 1+(j-1)d$ for j=1,…,M2 with $d=(b_2-1)/(M2-1)$ dividing $I_2=[1,b_2]$ uniformly $\Rightarrow a_i:= r_{max,i}^{-2}$ for i=M0+1,2,…M0+M2. Of course, certain orthogonal properties, etc. can also be used in the choice, as well as some $r_{max,i}$ can be in $I_1$ also, etc.. Figs. 1-2 show how Eqs.9-10 work.

More generally, the maximum of $y\equiv r^{2k}\exp(-ar^2)$ is at $r_{max}=(k/a)^{1/2}$, and the two inflexion points at r>0 is $r_{inf}=AK(i,k)$ where $A\equiv(2\sqrt{a})^{-1}$ and $K(i,k)\equiv$ sqrt[$(-1)^i$sqrt$(16k+1)+4k+1$] for i=1,2 and k=0,1,2…,, and the width (dist. between the two inflexion points) is $r_{inf,2}-r_{inf,1}= A(K(2,k)-K(1,k))$. The K(2,k)-K(1,k)= √2, 2.084, 2.0356, 2.0006, 2.0001 if k=0, 1, 2, 100, 10000, etc., i.e. a quasi-constant, so practically $r_{inf,2}-r_{inf,1}= (2a)^{-1/2}$ if k=0 and $\approx a^{-1/2}$ if k≥1 (as well as complex on 0<k<1/2). Importantly, the width depends as $a^{-1/2}$ and quasi-independent of k. Its consequence is that k>1 terms in Eq.8 may not necessary for an adequate fit, demonstrated on Figs. 1-2. This width corresponds to the "half width value" definition in practice, $r_2-r_1$, wherein $r_i$ are the two locations of $y_{max}/2$ values, i.e. the solution of $y_{max}/2\equiv (k/a)^k\exp(-k)/2= r_i^{2k}\exp(-ar_i^2)$, a transcendent equation for $r_2-r_1$ in contrast to the explicit $r_{inf,2}-r_{inf,1}= A(K(2,k)-K(1,k))$; furthermore, 0.95 (k=i=1) < $y(r_{inf,i})/(y_{max}/2)$ < 1.27 (k=1, i=2) for k≥1, tending below/above to 1.213 if k→∞ for i=1 or 2.

To obtain an effective and short series Gaussian expansion, the following cases have been considered:

**M point coincidence in interval (b1,b2)** for Eqs.9-10 uses e.g. the pre-determined set $\{a_i\}$ or $\{a_i\}\cup\{a,b\}$ above, and substitutes the M0 inflexion points $r_{inf,i}$ and M2 (Eq.9) or L+1 (Eq.10) maximums $r_{max,i}$ into $f(r_i)=|r_i|^{-u}$ with i=1,2,… M0+M2 or M0+L+1 which provides a LES for $\{c_i\}$ or $\{c_i\}\cup\{d_k\}$, resp..

**LSF for interval (b1,b2)** for Eqs.9-10 is supposed to be more accurate than the previous 'M point coincidence' and the parameter set $\{a_i\}$ or $\{a_i\}\cup\{a,b\}$ can be chosen in the same way. The LSF for parameter set $\{c_i\}$ or $\{c_i\}\cup\{d_k\}$ comes from standard LSF minimization of $Y\equiv \int_I[f(r)-|r|^{-u}]^2 dr$ by solving the LES from $\{\partial Y/\partial c_i=0\}$ and $\{\partial Y/\partial d_k=0\}$ yielding symmetric LSF matrix in $[LSF_{ij}][coeff.]=[B_i]$, where vector [coeff.]= $[c_1,…c_M]^T$ or $[c_1,…,c_{M0}, d_0, …d_L]^T$ is to be found. For Eq.9, $LSF_{ij}\equiv \int_{(b1,b2)} r^n \exp(-ar^2)dr$, where $a\equiv a_i+a_j$, its evaluation yields $(1/2)(\pi/a)^{1/2}$ (erf($b_2 a^{1/2}$)-erf($b_1 a^{1/2}$)) if n=0 at i,j≤M0. If n=2 at (i≤M0, j>M0) or (i>M0, j≤M0), if n=4 at i,j>M0, as well as for $B_i\equiv \int_{(b1,b2)} r^{n-u} \exp(-a_i r^2)dr$ with n= 0 at i≤M0 and n=2 at i>M0, and in case of Eq.10, for the LSF matrix elements $\int_{(b1,b2)}Mat(r)dr$, where $Mat(r)\equiv \exp(-(a_i+a_j)r^2)$, $\exp(-a_i r^2)(r^2-r_0^2)^j E(r)$ and $(r^2-r_0^2)^{i+j}(E(r))^2$ with $E(r)\equiv \exp(-ar^2)+B\exp(-br^2)$, as well as for LFS vector elements $\int_{(b1,b2)}Vec(r)dr$, where $Vec(r)\equiv r^{-u}\exp(-a_i r^2)$ and $r^{-u}(r^2-r_0^2)^k E(r)$ the primitive functions can be found in Appendix 1.

**Modified Taylor expansion at $r_0=1$ for interval (b1,b2)** for Eqs.9-10 is based on the standard Taylor expansion. An analytic function is an infinitely differentiable function such that the Taylor series at any point $r_0$ in its domain, $T(r)\equiv \Sigma_{n=0}^\infty f(r_0)^{(n)}(r-r_0)^n/n!$, converges to f(r) for r in a neighborhood of $r_0$ point-wise; its important property is that $T(r_0)^{(n)}= f(r_0)^{(n)}$ for the 0,1,2…$n^{th}$ derivatives, serving a base to perform the expansion (using that in $P(s)\equiv \Sigma_{k=0}^L d_k (s-s_0)^k$ the $P(s_0)^{(k)}/k!= d_k$); inexactly saying: "Two functions are close to each other in the neighborhood of $r_0$, if their first n derivatives are the same at $r_0$". Here we perform the expansion as $[d^n|r|^{-u}/dr^n]_{r0}:= [d^n f(r)/dr^n]_{r0}$ for n=0,1,…N for the derivatives at $r_0=1$, and the modification is that not pure polynomials, but the series in Eqs.9-10 are used.

First case is when we force equality for M derivatives at $r_0=1$ for the M parameters. Since $d^n\exp(-ar^2)/dr^n$ generates high degree polynomial multiplier (while $d^n s^k\exp(-as)/ds^n$ with k=0 or 1 does not), we make a trick as $s:= r^2$ ($\Rightarrow s_0=r_0^2=1$). From Eq.9, $|r|^{-u}=|s|^{-u/2}\approx \Sigma_{i=1}^{M0} c_i \exp(-a_i s)+ \Sigma_{i=M0+1}^{M0+M2} c_i s \exp(-a_i s) \Rightarrow [d^n|s|^{-u/2}/ds^n]_{s=1}= [\Sigma_{i=1}^{M0} c_i(-a_i)^n\exp(-a_i s) + \Sigma_{i=M0+1}^{M0+M2} (-a_i)^n (s-n/a_i) \exp(-a_i s)]_{s=1}$ for n=0,1…M-1, yielding a LES for $\{c_i\}$ with $LSF_{ji}\equiv (-a_i)^{j-1} (2+(1-j)/a_i) \exp(-a_i)$. However, the larger pre-determined $\{a_i\}$ for interval $(b_1,1)$ generates large values by $a_i^j$, for example, if M2=0 for simplicity, between LSF matrix elements (M,1) and (M,M) the ratio $(a_1/a_M)^{M-1} \exp(a_M-a_1)= (b_2/b_1)^{2(M-1)}\exp(0.5(b_2^{-2}-b_1^{-2}))$, which can cause instability in solving the LES. On the other hand, without using variable s in Eq.9, the $[d^n|r|^{-u}/dr^n]_{r=1}= [d^n f(r)/dr^n]_{r=1}$ can also be treated as LES to calculate $\{c_i\}$, using the elementary $d^n t^{-w}/dt^n= (-1)^n t^{-w-n} w (w+1) (w+2)…(w+n-1)$ for (t,w)= (s,u/2) or (r,u). Notice that, a good expansion with $s:= r^2$ e.g. for u=1 around $s=r^2=1$ for s in (0.01,25) guaranties for r the range $(\sqrt{0.01}, \sqrt{25})= (0.1, 5)$ only. The same procedure can be done with Eq.10 as well.

Second case is when we take advantage on polynomials in Eq.10 as in the Taylor series. With $r_0=1$, the $[d^n|s|^{-u/2}/ds^n]_{s=1}= [d^n f(s=r^2)/ds^n]_{s=1}$ leads to a LES for $\{c_i, d_k\}$, (or even a recursive formula to analytically evaluate if $M_0$ is small). Using $\Sigma\equiv\Sigma_{i=1}^{M0}$ and $e(i)\equiv (-a)^i\exp(-a)+ B(-b)^i\exp(-b)$, the $0^{th}$,…$n=6^{th}$ derivatives [9]

are $1= \Sigma c_i \exp(-a_i) + e(0)d_0$, $-u/2= -\Sigma a_i c_i \exp(-a_i) + e(1)d_0 + e(0)d_1$, $u(u+2)/4 = \Sigma a_i^2 c_i \exp(-a_i) + e(2)d_0 + 2e(1)d_1 + 2e(0)d_2$, $-u(u+2)(u+4)/8 = -\Sigma a_i^3 c_i \exp(-a_i) + e(3)d_0 + 3e(2)d_1 + 6e(1)d_2 + 6e(0)d_3$, as well as the right sides are $f^{(4)}(s=r^2=1) = \Sigma a_i^4 c_i \exp(-a_i) + e(4)d_0 + 4e(3)d_1 + 12e(2)d_2 + 24e(1)d_3 + 24e(0)d_4$, $f^{(5)}(s=1) = -\Sigma a_i^5 c_i \exp(-a_i) + e(5)d_0 + 5e(4)d_1 + 20e(3)d_2 + 60e(2)d_3 + 120e(1)d_4 + 120e(0)d_5$, $f^{(6)}(s=1) = \Sigma a_i^6 c_i \exp(-a_i) + e(6)d_0 + 6e(5)d_1 + 30e(4)d_2 + 120e(3)d_3 + 360e(2)d_4 + 720e(1)d_5 + 720e(0)d_6$. The first $0,\ldots,n^{th}$ derivatives provides n+1 equations and coefficients $(r^2-r_0^2)^k$ at r=1 eliminate $d_k$ with k>n (like in basic Taylor expansion). There are $M_0$ terms with $c_i$ in all n≥ derivatives, so pivot points can be picked up from $(b_1,b_2)$ for the LES, which is an advantage along with that $a_i^n$ comes up with lower n (compare to previous case). The benefit of e(i) as a function of {a,B,b} is that one can use pre-determined values to tune, but better if one can force to pivot even extra three points to coincide with function values, e.g. the two $f(s=r^2=b_i^2)= b_i^{-u}$ for i=1,2 beside $f(s=r^2=1)=1^{-u}=1$, however, it is to solve a transcendent equation system for {a,B,b}, so one must iterate from an initial value from 1= f(1). Simplification can be made with B:=0 for Eq.10. This pivoting is an advantage of this modified Taylor expansion over the standard Taylor expansion. Notice that, in standard Taylor expansion the coefficients decrease by the division with n!, but here $|d_k|$ increases with k counterbalanced with $\exp(-r^2)$.

Without the $s=r^2$ substitution in Eq.10 with $r_0=1$, the direct derivatives $[d^n|r|^{-u}/dr^n]_{r=1} = [d^n f(r)/dr^n]_{r=1}$ leads [9] analogously to a bit more complicated expressions such as $f(r=1)= \Sigma c_i \exp(-a_i) + e(0)d_0$, $f^{(1)}(r=1)= \Sigma(-2a_i)c_i\exp(-a_i) + 2e(1)d_0 + 2e(0)d_1$, $f^{(2)}(r=1)= \Sigma(2a_i(2a_i-1))c_i\exp(-a_i) + [4e(2)+ 2e(1)]d_0 + [8e(1)+ 2e(0)]d_1 + 8e(0)d_2$, $f^{(3)}(r=1)= \Sigma(-4a_i^2(2a_i-3))c_i\exp(-a_i) + [8e(3)+ 12e(2)]d_0 + [24e(2)+ 24e(1)]d_1 + [48e(1)+ 24e(0)]d_2 + 48e(0)d_3$, $f^{(4)}(r=1)= \Sigma(4a_i^2(4a_i^2-12a_i+3))c_i\exp(-a_i) + [16e(4)+ 48e(3)+ 12e(2)]d_0 + [64e(3)+ 144e(2)+ 24e(1)]d_1 + [192e(2)+ 288e(1)+ 24e(0)]d_2 + [384e(1)+ 288e(0)]d_3 + 384e(0)d_4$, etc.. The LES (or recursive) treatment is available again, so the pivoting.

**A good fit for $1/r^u$:** Only one fit for the case u=1 is exhibited with LSF for Eq.9 choosing interval $(b_1,b_2)= (0.04, 50.0)$, $M=M_0+M_2=50$, $M_0=33$, $M_2=17$. <u>For the main goal in title, a good fit is fundamental!</u> By the extreme decrease of $1/r^u$ with u>0 and r in (0,1] and flatness in [1,∞), a steep correction $C1 \exp(-A1\, r^2) + C2 \exp(-A2\, r^2)$ was added to improve r in (0.001,0.04) with (C1,A1,C2,A2)= (1340.0, 500000.0, 155.0, 5000.0) or in (0.01,0.04) with (96.0, 5000.0, 10.5, 1250.0), both has minor contribution in our main task, generally in $\int_{(0,\infty)}(1/r^u)\exp(-a(r-r_0)^2)dr= 4\pi\int_{(0,\infty)}r^{2-u}\exp(-a(r-r_0)^2)dr$, in which this correction is in effect in part $\int_{(0,0.04)}$ only. Recall the $\Gamma(z)$ and $\gamma(z,x)$ in Appendix 1 for which u the $\int_{(0,x)}=\infty$ happen! In this way, the $M_0$ is effectively 35, not 33. The area under the curve of the left side in Eq.9 is $\ln(b_2/b_1)= 7.1308$ vs. the fit for right side 7.1303, as well as the function value deviation is less than 0.05%. The M pair $(a_i,c_i)$ values are listed in Appendix 2 and plotted on Fig.3.

## Evaluation of generalized Coulomb integrals with Gaussian expansion

The evaluation in Eqs.3-7 and related cases are based [2, 4] on Laplace transformation with $\exp(-a^2 t^k)$ in integrand with k=1,2, but more importantly, the way of evaluation depends somehow on the restricted integer (n,m)= (1,0) or (2,0) values in W. In detail, $a^{-1}= \pi^{-1/2}\int_{(-\infty,\infty)}\exp(-a^2 t^2)dt$ and $a^{-2} = \int_{(0,\infty)} \exp(-a^2 t)dt$ are used for $a:= R_{C1}$ or $r_{12}$ to transform $a^{-(1\text{ or }2)} \to \exp(-a^2(\ldots))$ in the integrand, and an extra inner integral with t comes up. This trick is replaced with using Eqs.8-10 for $r:= R_{C1}$ or $r_{12}$, and this $t\to \{a_i\}$ buys off the integration over $a_i$ on the price of having (L+1)M terms in LC instead of one. As a result, in our evaluation of Eq.1, the algorithm/procedure is basically the same scheme and works for any real (n,m) in W. Since ρ in Eq.1 is LC of Gaussians in Eq.2, Eq.1 breaks up into LC of $\int G_{P1}(p,nx1,ny1,nz1)\, G_{Q2}(q,nx2,ny2,nz2)\, G_{S3}(s,nx3,ny3,nz3)\, W(\mathbf{r}_1,\mathbf{r}_2,\mathbf{r}_3)\, d\mathbf{r}_1 d\mathbf{r}_2 d\mathbf{r}_3$. Applying Eq.9, the 1s-like $\exp(-a_i r^2)$ and 3d-like $r^2\exp(-a_i r^2)$ terms replace $r^{-u}$, but for the sake of brevity, we discuss the 1s-like cases below, because the exponential part is problematic only.

**Cases W(1)= $R_{C1}^{-n}$ and $R_{C1}^{-n}R_{D1}^{-m}$** need the fit for u= n and for both, n and m, resp. for $\int G_{P1}(p,nx1,ny1,nz1)W(\mathbf{r}_1)d\mathbf{r}_1$. With $r:= R_{C1}$ in Eq.9, a good LSF fit provides a LC with parameter set $\{c_i,a_i\}$, and the integral becomes a LC of $\int G_{P1}(p,nx,ny,nz)\, R_{C1}^{2k}\exp(-a_i R_{C1}^2)d\mathbf{r}_1 = \int (x_1-R_{Px})^{nx}(y_1-R_{Py})^{ny}(z_1-R_{Pz})^{nz} R_{C1}^{2k}\exp(-pR_{P1}^2-a_i R_{C1}^2)d\mathbf{r}_1$ with k= 0,1 for which the exponent can be shifted to a common point (Appendix 3), thereafter the power terms can be re-centered to this common point T (Appendix 4), finally Appendix 1 provides the analytic evaluation. In case of k=0 and 1s type Gaussian (wherein nx=ny=nz=0 and $\mathbf{R}_T \equiv (p\mathbf{R}_P + a_i\mathbf{R}_C)/(p+a_i)$ is irrelevant), $\int\exp(-pR_{P1}^2-a_i R_{C1}^2)d\mathbf{r}_1 = \int \exp(-(p+a_i)R_{T1}^2 - pa_i R_{PC}^2/(p+a_i))d\mathbf{r}_1 = \exp(-pa_i R_{PC}^2/(p+a_i))\int \exp(-(p+a_i)R_{T1}^2)d\mathbf{r}_1$, and

$$\int \exp(-pR_{P1}^2 - a_i R_{C1}^2)\, d\mathbf{r}_1 = (\pi/(p+a_i))^{3/2}\exp(-pa_i R_{PC}^2/(p+a_i)). \tag{11}$$

Its proper LC is an alternative expression to Eqs.3-4 wherein n=1 or 2 only, but now n can be any real number.

If $W= R_{C1}^{-n}R_{D1}^{-m}$, one must take both sets, $\{c^n_i\}$ and $\{c^m_i\}$, from LSF, where the upper index indicates that these belongs to $R_{C1}^{-n}$ and $R_{D1}^{-m}$, resp., (the set $\{a_i\}$ and the value of M can be common), so $W= R_{C1}^{-n} R_{D1}^{-m} \approx$ product of terms in Eq.9 with $r:=R_{C1}$ and $r:= R_{D1}$. This is a LC of terms $R_{C1}^{2kn} R_{D1}^{2km} \exp(-a_i R_{C1}^2 - a_j R_{D1}^2)= R_{C1}^{-2kn} R_{D1}^{-2km} \exp(-(a_i+a_j)R_{T1}^2) \exp(-a_i a_j R_{CD}^2/(a_i+a_j))$, with kn, km=0,1 using the common points (Appendix 3) $\mathbf{R}_T \equiv (a_i\mathbf{R}_C + a_j\mathbf{R}_D)/(a_i+a_j)$ for the $M^2$ terms. Most importantly, algebraically W is a LC just like in the previous case $W= R_{C1}^{-n}$. It means that the algorithm is also the same, and the simplest 1s case (nx=ny=nz=0) with kn=km=0 also leads essentially to the same expression as in the right of Eq.11 as
$(\pi/(p+a_i+a_j))^{3/2} \exp(-p(a_i+a_j)R_{PT}^2/(p+a_i+a_j))$.

**Case W(1,2)=** $r_{12}^{-n}$ for $\int G_{P1}(p,nx1,ny1,nz1) G_{Q2}(q,nx2,ny2,nz2) W(\mathbf{r}_1,\mathbf{r}_2) d\mathbf{r}_1 d\mathbf{r}_2$ needs the fit for u= n to use $r:= r_{12}^{-n}$ in Eq.9 with the set $\{c_i,a_i\}$ from LSF, and the integral becomes a LC of $\int G_{P1}(p,nx1,ny1,nz1) G_{Q2}(q,nx2,ny2,nz2) r_{12}^{2k} \exp(-a_i r_{12}^2) d\mathbf{r}_1 d\mathbf{r}_2 = \int (x_1-R_{Px})^{nx1} (x_2-R_{Qx})^{nx2} (x_1-x_2)^{2k} (\ldots) \exp(-(pR_{P1}^2 + qR_{Q2}^2 + a_i r_{12}^2)) d\mathbf{r}_1 d\mathbf{r}_2$ with k=0,1. In case of k=0 and 1s type Gaussians (nx1=…=nz2=0), one can apply E.9 for $\mathbf{r}_1$ with $R_C \to \mathbf{r}_2$ as $\int \exp(-(pR_{P1}^2 + qR_{Q2}^2 + a_i r_{12}^2)) d\mathbf{r}_1 d\mathbf{r}_2 = \int [\int \exp(-pR_{P1}^2 - a_i r_{12}^2)) d\mathbf{r}_1] \exp(-qR_{Q2}^2) d\mathbf{r}_2 = (\pi/(p+a_i))^{3/2} \int \exp(-pa_i R_{P2}^2/(p+a_i)) \exp(-qR_{Q2}^2) d\mathbf{r}_2$ for which, again, the exponent can be shifted to a common point using Appendix 3, thereafter the power terms can be re-centered to this common point T using Appendix 4, finally Appendix 1 can used to evaluate (the location of T is irrelevant in case of k=0 and 1s), or more elegantly, re-use Eq.11 with $1 \to 2$, $p \to pa_i/(p+a_i)$, $a_i \to q$ strictly for the integral obtaining
$$\int \exp(-pR_{P1}^2 - qR_{Q2}^2 - a_i r_{12}^2) d\mathbf{r}_1 d\mathbf{r}_2 = \pi^3 (qp+qa_i+pa_i)^{-3/2} \exp(-pqa_i/(qp+qa_i+pa_i)R_{PC}^2), \quad (12)$$
and its proper LC is an alternative to Eqs.5-6 in which n=1, 2, but Eq.12 is more general, i.e. n can be any real number.

For higher power than 1s (i.e. when the order of polynomial in Eq.2 is higher than zero) there is another standard way beside the straightforward generalization of Eq.12: The $\int (x_1-R_{Px})^{nx1} (x_2-R_{Qx})^{nx2} (x_1-x_2)^{2k} (\ldots) \exp(-(pR_{P1}^2 + qR_{Q2}^2 + a_i r_{12}^2)) d\mathbf{r}_1 d\mathbf{r}_2 = \int (x_1-R_{Px})^{nx1}(x_2-R_{Qx})^{nx2} (x_1-x_2)^{2k} \exp(-(p(x_1-R_{Px})^2 + q(x_2-R_{Qx})^2 + a_i(x_1-x_2)^2)) dx_1 dx_2 \int(\ldots) dy_1 dy_2 \int (\ldots) dz_1 dz_2$ separation shows three algebraically equivalent terms in the product. The basic algebraic difficulty comes from that in the exponential the variables are not separated as sums, i.e. there are bad mixed products: Although no $x_1y_2$, etc., but we must get rid of terms $x_1x_2$, $y_1y_2$ and $z_1z_2$. Each can be treated analogously by using the procedure in Appendix 5.

**Case W(1,2)=** $R_{C1}^{-n} r_{12}^{-m}$ for $\int G_{P1}(p,nx1,ny1,nz1) G_{Q2}(q,nx2,ny2,nz2) W(\mathbf{r}_1,\mathbf{r}_2) d\mathbf{r}_1 d\mathbf{r}_2$ needs a LSF fit for u= n and m to obtain the sets $\{c^n_i\}$ and $\{c^m_i\}$, resp., where the upper index indicates that these belong to $R_{C1}^{-n}$ and $r_{12}^{-m}$, resp., (the set $\{a_i\}$ and the value of M can be common). $W = R_{C1}^{-n} r_{12}^{-m} \approx$ product of terms in Eq.9 with $r:= R_{C1}$ and $r:= r_{12}$ which is a LC of terms $c^n_j c^m_i R_{C1}^{2kn} r_{12}^{2km} \exp(-a_j R_{C1}^2 - a_i r_{12}^2)$ with kn,km=0,1, so the integral becomes a LC of $\int G_{P1} G_{Q2} R_{C1}^{2kn} r_{12}^{2km} \exp(-a_j R_{C1}^2 - a_i r_{12}^2) d\mathbf{r}_1 d\mathbf{r}_2 =$
$$\int P(x_1,x_2)P(y_1,y_2)P(z_1,z_2) \exp(-(pR_{P1}^2+qR_{Q2}^2+a_j R_{C1}^2+a_i r_{12}^2)) d\mathbf{r}_1 d\mathbf{r}_2 =$$
$$\exp(-a_j p R_{CP}^2/(a_j+p)) \int P(x_1,x_2)P(y_1,y_2)P(z_1,z_2) \exp(-((a_j+p)R_{T1}^2 + qR_{Q2}^2 + a_i r_{12}^2)) d\mathbf{r}_1 d\mathbf{r}_2 \quad (13)$$
where $P(w_1,w_2) \equiv (w_1-R_{Pw})^{nw1} (w_2-R_{Qw})^{nw2} (w_1-R_{Cw})^{2kn}(w_1-w_2)^{2km}$ for w=x,y,z with the common point $\mathbf{R}_T \equiv (a_j\mathbf{R}_C + p\mathbf{R}_P)/(a_j+p)$ via Appendix 3, (or just treat the exponent as $(p(x_1-R_{Px})^2 + q(x_2-R_{Qx})^2 + a_j(x_1-R_{Cx})^2 + a_i(x_1-x_2)^2) + (.) + (.)$). In case of kn=km=0 and 1s type Gaussian (nx1=…=nz2=0) this case falls into the form of Eq.12 in relation to integral evaluation, and the procedure for higher power than 1s (i.e. when the order of polynomial in Eq.2 is higher than zero) the way to get rid of terms $x_1x_2$, $y_1y_2$ and $z_1z_2$ can be found again in Appendix 5.

**Case W(1,2,3)=** $r_{12}^{-n} r_{13}^{-m}$ for $\int G_{P1}(p,nx1,ny1,nz1) G_{Q2}(q,nx2,ny2,nz2) G_{S3}(s,nx3,ny3,nz3) W(\mathbf{r}_1,\mathbf{r}_2,\mathbf{r}_3) d\mathbf{r}_1 d\mathbf{r}_2 d\mathbf{r}_3$ needs a LSF fit for u= n and m to obtain the sets $\{c^n_i\}$ and $\{c^m_i\}$, resp., where the upper index indicates that these belong to $r_{12}^{-n}$ and $r_{13}^{-m}$, resp., (the set $\{a_i\}$ and the value of M can be common). $W = r_{12}^{-n} r_{13}^{-m} \approx$ product of terms in Eq.9 with $r:= r_{12}$ and $r:= r_{13}$ which is a LC of terms $c^n_i c^m_j r_{12}^{2kn} r_{13}^{2km} \exp(-a_i r_{12}^2 - a_j r_{13}^2)$ with kn,km=0,1, so the integral becomes a LC of $\int G_{P1} G_{Q2} G_{S3} r_{12}^{2kn} r_{13}^{2km} \exp(-a_i r_{12}^2 - a_j r_{13}^2) d\mathbf{r}_1 d\mathbf{r}_2 d\mathbf{r}_3 = \int (x_1-R_{Px})^{nx1}(x_2-R_{Qx})^{nx2}(x_3-R_{Sx})^{nx3} (x_1-x_2)^{2kn}(x_1-x_3)^{2km} (\ldots) \exp(-(pR_{P1}^2 + qR_{Q2}^2 + sR_{S3}^2 + a_i r_{12}^2 + a_j r_{13}^2)) d\mathbf{r}_1 d\mathbf{r}_2 d\mathbf{r}_3$. The separation of exponent as $(p(x_1-R_{Px})^2 + q(x_2-R_{Qx})^2 + s(x_3-R_{Sx})^2 + a_i(x_1-x_2)^2 + a_j(x_1-x_3)^2) + (.) + (.)$ yield products of three $\int(\ldots) d\mathbf{r}_i$. The procedure for 1s type Gaussians (nx1=…=nz3=0) and higher than 1s (i.e. when the order of polynomial in Eq.2 is higher than zero) the way to get rid of terms $x_1x_2$, $y_1y_2$ and $z_1z_2$ can be found again in Appendix 5, but now in three dimension, it is more convenient to obtain the eigenvalues and eigenvectors by standard calculation to build up the matrix O, instead of using the three dimension rotation matrix. This case is an alternative to Eq.7.

From the cases detailed above, it is quite obvious that the scheme introduced is more unified and simpler, more importantly opens a broader range for (n,m) in W in the applications than the solution schemes for Eqs.3-7 restricted to n=m=0,1,2.

**Appendix 1:** The $erf(x) \equiv \pi^{-1/2} \int_{(-x,x)} exp(-w^2)dw = 2\pi^{-1/2} \int_{(0,x)} exp(-w^2)dw$, for which $erf(\infty)=1$ as well as $\int_{(b_1,b_2)} exp(-aw^2)dw = (1/2)(\pi/a)^{1/2} (erf(b_2 a^{1/2}) - erf(b_1 a^{1/2}))$; erf(x) is also a FORTRAN command for real x.

The gamma function is $\Gamma(z) \equiv \int_{(0,\infty)} w^{z-1} exp(-w)dw = 2\int_{(0,\infty)} w^{2z-1} exp(-w^2)dw$. (It follows that the maximum of the two integrands provides (via $\int_{(0,\infty)} f(w)dw \approx max(f)$) a weaker but similar formulas to Stirling's approx. as $n! \approx (n/e)^n$ and $2((n+0.5)/e)^{n+0.5}$. Alternatively, $\Gamma(z) \equiv \int_{(0,1)} (ln(1/w))^{z-1} dw$.) The gamma(z) is a FORTRAN command to calculate $\Gamma(z)$ for real z except at $\Gamma(z=0,-1,-2,...)= \pm\infty$. Algorithms restrict to e.g. $0<z\leq 4$ with shift $sin(\pi z)\Gamma(z)\Gamma(1-z)=\pi$ and recursive forms $\Gamma(z+1)=z\Gamma(z)$, $\Gamma(z+k)= \Gamma(z) \Pi_{i=0}^{k-1}(z+i) \equiv \Gamma(z)(z+k-1)!$, etc.. The "upper incomplete gamma function" is defined as $\Gamma(z,x) \equiv \int_{(x,\infty)} w^{z-1} exp(-w)dw$ and the "lower" one $\gamma(z,x)$ is with domain (0,x). Well known extension of $\Gamma(z)$ is from real z to complex z in math, however another extension is in CC as $\Gamma_3(z) \equiv \int_{(0,\infty)} r^{z-1} exp(-r)d\mathbf{r} = 4\pi\int_{(0,\infty)} r^{z+1} exp(-r)dr = 4\pi\Gamma(z+2)$, and a consequence (of this simple relation) is that while $\Gamma(z=0 \text{ or } -1) = \pm\infty$, the $\Gamma_3(z=0 \text{ or } -1)$ = finite, and so on. $\Gamma_3(z)$ comes up in Coulomb potentials (z=0) of molecules and corrections ($z\neq 0$), see main title.

Stable convergences are provided by the algorithms: For x>1 the $\Gamma(z,x) \approx exp(-x)x^z/h(z,x)$, where the $k^{th}$ depth of continued fraction h(z,x) can be approximated as p(k)/q(k) via y(k+2)= xy(k+1)+ (1+k/2)y(k) for even k and y(k+2)= y(k+1)+ (1.5–z+k/2)y(k) for odd k with starting values p(0)=x, p(1)=x+1–z, q(0)= q(1)= 1. For $0\leq x\leq 1$ the $\gamma(z,x) \equiv \int_{(0,x)} w^{z-1} exp(-w)dw = x^z \Sigma_{k=0}^{\infty} (-x)^k/(k!(z+k))$ wherein the everywhere converging power series for exp(-w) was used. These two algorithms can be connected by knowing $\Gamma(z,x) + \gamma(z,x) = \Gamma(z)$, $\gamma(z=0,-1,-2,...,x>0)= \pm\infty$ and $\Gamma(z,x>0)$=finite; unfortunately e.g. FORTRAN does not support algorithm neither for $\Gamma(z,x)$ nor for $\gamma(z,x)$.

For m= 1 and 2, the $\int_{(0,\infty)} x^n exp(-ax^m)dx = \Gamma[(n+1)/m]/(m\ a^{(n+1)/m})$ holds for a>0. If m=2 and n=0 $\Rightarrow$ $\int_{(R3)} exp(-ar_1^2)d\mathbf{r}_1 = (\int_{(-\infty,\infty)} exp(-ax_1^2)dx_1)^3 = (\pi/a)^{3/2}$. If m=2 $\Rightarrow \int_{(-\infty,\infty)} x^n exp(-ax^2)dx = \Gamma[(n+1)/2]/a^{(n+1)/2}$ for even n, but zero if n is odd. The $\Gamma[n+1]= n!$ for n=0,1,2,..., with $\Gamma[n+1/2]= 1\times 3\times 5\times ...(2n-1)\ \pi^{1/2}/2^n$ for n=1,2,... and $\Gamma[1/2]= \pi^{1/2}$.

For $\int_{(b_1,b_2)} r^{n-u} exp(-ar^2)dr$ in LSF for Eqs.9-10, the general indefinite integral [9] with real (a,v) is $I(a,v) \equiv \int exp(-ax^2)/x^v dx = (-1/2)x^{1-v}(ax^2)^{(v-1)/2}\Gamma((1-v)/2,ax^2) + const.$, which is $(-1/2)a^{(v-1)/2}\Gamma((1-v)/2,ax^2) + const.$ for a,v,x>0. The primitive function (w/o const.) simplifies to 1., use of "real exponential integral" $Ei(x) \equiv -\int_{(-x,\infty)} (exp(-w)/w)dw$ for odd v, e.g. I(a,1)= $Ei(-ax^2)/2$, etc., 2., use of erf(x) for even v, e.g. -I(a,2)= $(\pi a)^{1/2} erf(a^{1/2}x) + exp(-ax^2)/x$, etc., 3., otherwise the $\Gamma$ must be used, e.g. -I(a,1/2)= $\Gamma(1/4,ax^2)/(2a^{1/4})$, I(a,3/2)= $2a^{1/4}\Gamma(0.75,ax^2) - x^{-1/2}exp(-ax^2)$, etc.. At v=1 the algorithm for $\Gamma(0,0<ax^2<1)$ has problem, there I(a,v=1)= $Ei(-ax^2)/2$. One must not forget that if $\Gamma(z,x)$ or Ei(x) are not available in a program language, the integral $\int_{(b_1,b_2)} r^{n-u} exp(-ar^2)dr$ can still be conveniently calculated directly using simpler standard numerical integral methods for $b_1>0$ and not a very large $b_2$.

**Appendix 2:** For i=1...M0, $a_i$= 312.5, 102.040816, 50.0, 29.585799, 19.531250, 13.850416, 10.330579, 8.0, 6.377551, 5.202914, 4.325260, 3.652301, 3.125000, 2.704164, 2.362949, 2.082466, 1.849112, 1.652893, 1.486326, 1.343725, 1.220703, 1.113834, 1.020408, 0.938262, 0.865651, 0.801154, 0.743605, 0.692042, 0.645661, 0.603792, 0.565867, 0.531406, 0.5. For i=M0+1...M, $a_i$= 0.112697, 0.028556, 0.012749, 0.007188, 0.004606, 0.003202, 0.002354, 0.001803, 0.001425, 0.001155, 0.000955, 0.000802, 0.000684, 0.000590, 0.000514, 0.000452, 0.000400. For i=1...M0, $c_i$= 0.267712E+02, -0.373905E+02, 0.247249E+03, -0.147319E+04, 0.800623E+04, -0.369627E+05, 0.141068E+06, -0.437026E+06, 0.108172E+07, -0.209903E+07, 0.310550E+07, -0.333751E+07, 0.233348E+07, -0.645937E+06, -0.557682E+06, 0.580941E+06, 0.429097E+06, -0.131691E+07, 0.321697E+05, 0.340478E+07, -0.475887E+07, 0.144323E+07, 0.220207E+07, -0.189648E+07, -0.133979E+05, 0.868564E+05, 0.132827E+07, -0.218205E+07, 0.960134E+06, 0.141437E+07, -0.225072E+07, 0.122850E+07, -0.246353E+06. For i=M0+1...M, $c_i$= 0.742357E-01, 0.675994E-02, -0.217629E-03, 0.335532E-02, -0.903693E-02, 0.258367E-01, -0.525187E-01, 0.564849E-01, 0.233833E-01, -0.172877E+00, 0.249763E+00, -0.173240E+00, 0.560859E-01, -0.596661E-02, 0.108651E-03, 0.532194E-04, 0.293131E-04.

**Appendix 3**: The product of two Gaussians, $G_{J1}(p_J,0,0,0)$ with J=1,...,m=2 is another Gaussian centered somewhere on the line connecting the original Gaussians, but a more general expression for m>2 comes from the elementary $\Sigma_J p_J R_{J1}^2 = (\Sigma_J p_J) R_{T1}^2 + (\Sigma_J \Sigma_K p_J p_K R_{JK}^2)/(2\Sigma_J p_J)$ and $\mathbf{R}_T \equiv (\Sigma_J p_J \mathbf{R}_J)/(\Sigma_J p_J)$, where $\Sigma_{J \text{ or } K} \equiv \Sigma_{(J \text{ or } K=1)}^m$ and $R_{J1} \equiv |\mathbf{R}_J - \mathbf{r}_1|$ for $exp(\Sigma_J c_J) = \Pi_{(J=1)}^m exp(c_J)$, keeping in mind that $R_{JJ}=0$, and the m centers do not have to be collinear. For m=2, this reduces to

$$p R_{P1}^2 + q R_{Q1}^2 = (p+q) R_{T1}^2 + pqR_{PQ}^2/(p+q) \tag{14}$$

yielding the well known and widely used

$$G_{P1}(p,0,0,0)\ G_{Q1}(q,0,0,0) = G_{T1}(p+q,0,0,0)exp(-pqR_{pq}^2/(p+q))\ . \tag{15}$$

We also need the case m=3, which explicitly reads as

$$p R_{P1}^2 + q R_{Q1}^2 + s R_{S1}^2 = (p+q+s) R_{T1}^2 + (pqR_{PQ}^2+psR_{PS}^2+qsR_{QS}^2)/(p+q+s) . \quad (16)$$

Only the $G_{T1}(p+q+s,0,0,0)$ depends on electron coordinate $\mathbf{r}_1$ not its multiplier, indicating that the product of Gaussians decomposes to (sum of) individual Gaussians, (s=0 reduces Eq.16 to 14).

**Appendix 4:** Given a single power term polynomial at $\mathbf{R}_P$, like in Eq.2, we need to rearrange or shift it to a given point $\mathbf{R}_S$. For variable x, this rearrangement is $(x-x_P)^n = \Sigma_{i=0}^n c_i(x-x_S)^i$, which can be solved systematically and directly for $c_i$ by the consecutive equation system obtained from the 0,1,…n$^{th}$ derivative of both sides at x:= $x_S$, yielding $POLY(x,P,S,n) \equiv (x-x_P)^n = \Sigma_{i=0}^n (^n_i)(x_S-x_P)^{n-i}(x-x_S)^i$, where $(^n_i)=n!/(i!(n-i)!)$. It reduces to the simpler well known binomial formula as $(x-x_P)^n = \Sigma_{i=0}^n (^n_i)(-x_P)^{n-i}x^i$ if $x_S=0$. Applying it two times yields the coefficients $\{C_i\}$ for $(x-x_P)^{n1}(x-x_Q)^{n2} = \Sigma_{i=0}^{n1+n2} C_i (x-x_S)^i$.

**Appendix 5:** Elementary derivation for variable $x \equiv (x_1,…x_n)^T$ provides [10] the known expression

$$\int_{(-\infty,\infty)}…\int_{(-\infty,\infty)} \exp(-x^THx/2+Jx) \, dx_1…dx_n = ((2\pi)^n /\det H)^{1/2}\exp(J H^{-1} J/2) \quad (17)$$

where H is a real symmetric matrix ($\Rightarrow$ owns real eigenvalues) and vector J is column or row vector, accordingly. Eq.17 with n=2 can be used directly for 1s type Gaussians $g_{Ai}$ e.g. in Eq.12. If polynomial factor is higher than unity in Eq.2, the Eq.17 with n=2 is not enough for Gaussians $G_{Ai}$ e.g. in Eq.13, one must get rid of the term $\exp(x_1x_2)$ in other way: Orthogonal or unitary transformation is needed, what helps in two dimensions (for $dx_1dx_2$) as follows: The exponent e.g. in Eq.12 contains $p(x_1-R_{Px})^2 +q(x_2-R_{Qx})^2 +a_i(x_1-x_2)^2 = (p+a_i)x_1^2 +(q+a_i)x_2^2 -2a_ix_1x_2 -2pR_{Px}x_1 -2qR_{Qx}x_2 +(pR_{Px}^2+qR_{Qx}^2)$, from which $\exp(-pR_{Px}^2-qR_{Qx}^2)$ can be pulled out from integrand, and $h_{11} \equiv -2(p+a_i)$, $h_{22} \equiv -2(q+a_i)$, $h_{12}=h_{21} \equiv 2a_i$ showing that the symmetric property is always provided for H in Eq.12, as well as $j_1 \equiv -2pR_{Px}$ and $j_2 \equiv -2qR_{Qx}$; notice that $a_i \neq 0$, so $x_1x_2$ is always present. H is a real symmetric matrix, so we can choose a square O to be orthogonal (columns and rows are $\perp$, i.e. $O^TO= OO^T= I$), and hence also a unitary matrix (conjugate transpose is also its inverse, i.e. $O^*O= OO^*= I$). We choose O such that $O^THO$ is diagonal: $x^THx= (x^TO)(O^THO)(O^Tx)$. O is a rotation matrix with upper row (cos t, –sin t) and lower (sin t, cos t) as well as $O^T= O^{-1}$. Using substitution x=Ou, i.e. $x_1=u_1\cos t-u_2\sin t$ with $x_2=u_1\sin t+u_2\cos t$, the $h_{11}x_1^2+ h_{22}x_2^2+ 2h_{12}x_1x_2= h_{11}(u_1\cos t-u_2\sin t)^2+ h_{22}(u_1\sin t+u_2\cos t)^2+ 2h_{12}(u_1\cos t-u_2\sin t) (u_1\sin t+u_2\cos t)= (.)u_1^2+ (.)u_2^2+ [2(h_{22}-h_{11})\cos t \sin t+ 2h_{12}(\cos^2t-\sin^2t)]u_1u_2$. The trick is that rotation angle t must be chosen to zero out the term $u_1u_2$, that is, $(h_{22}-h_{11})\cos t \sin t= h_{12}(2\sin^2t-1)$, that is (p-q) $(1-\sin^2t)^{1/2} \sin t= a_i(1-2\sin^2t)$ which is a 2$^{nd}$ order equation for sin t. Its solution is $\sin^2t= (1 \pm b/(4+b^2)^{1/2})/2 < 1$ with $b \equiv (p-q)/a_i$, always yielding a real valued angle t. Alternatively, one can find the eigenvectors (for columns of O) with standard way. The two real eigenvalues are solutions of the characteristic polynomial of symmetric H, i.e. one has to solve the $(h_{11}-\lambda)(h_{22}-\lambda) -h_{12}^2= 0$. Since the coordinate transformation is simply a rotation, the Jacobian determinant of the transformation yields $dx_1dx_2=du_1du_2$ and

$$\int_{(-\infty,\infty)}\int_{(-\infty,\infty)} P(x_1,x_2) \exp(-x^THx/2) \, dx_1dx_2 = \int_{(-\infty,\infty)}\int_{(-\infty,\infty)} Q(u_1,u_2) \exp(-(\lambda_1u_1^2+ \lambda_2u_2^2)/2) \, du_1du_2 \quad (18)$$

along with $Jx= j_1x_1+j_2x_2= j_1(u_1\cos t-u_2\sin t)+j_2(u_1\sin t+u_2\cos t)$. Finally, no $u_1u_2$ cross term in the integrand in Eq.18. This shows how one can prove Eq.17 as well, but more importantly, the polynomial part can be treated as $P(x_1,x_2) \equiv (x_1-R_{Px})^{n1}(x_2-R_{Qx})^{n2}= (u_1\cos t-u_2\sin t-R_{Px})^{n1}(u_1\sin t+u_2\cos t-R_{Qx})^{n2}= \Sigma c_{nm}u_1^nu_2^m \equiv Q(u_1,u_2)$ and finally $\int (x_1-R_{Px})^{n1}(x_2-R_{Qx})^{n2} \exp(-(p(x_1-R_{Px})^2 +q(x_2-R_{Qx})^2 +a_i(x_1-x_2)^2))dx_1dx_2$ breaks into the LC of product of $\int u_1^n \exp(-\lambda_1u_1^2/2+ \text{const.}u_1)du_1$ elementary, one dimension integrals in Appendix 1.

## ACKNOWLEDGMENTS

Financial and emotional support for this research from OTKA-K 2015-115733 and 2016-119358 are kindly acknowledged. Special thanks to Szeger Hermin for her help in typing the manuscript. The subject has been presented in ICNAAM_2019, Greece, Rhodes and AIP 2020.

## REFERENCES

1.: A.Szabo, N.S.Ostlund: Modern Quant. Chem.: Intro. Adv. Electr. Struct. Theory, 1982, McMillan, NY.
2.: S.Kristyan, AIP Conference Proceedings 1978, 470030-1 to 6 (2018),
   see also https://arxiv.org/ and https://chemrxiv.org for kristyan.
3.: W.Klopper, F.R.Manby, S.Ten-No, E.F.Valeev, Int. Rev. in Physical Chemistry, **25**, 427–468 (2006).
4.: S.Reine, T.Helgaker, R.Lindh, WIREs Comput. Mol. Sci. **2**, 290-303 (2012).
5.: S.Kristyan, Computers in Physics **8**, 556-575 (1994).
6.: P.M.W.Gill, Advances in Quantum Chemistry, 25,141-205 (1994).
7.: G.Samu, M.Kállay, J. of Chemical Physics, **146**, 204101 (2017).
8.: D.Kolb, R.Y.Cusson, Zeitschrift für Physik, **253**, 282–288 (1972).
9.: https://www.wolframalpha.com/calculators/integral-calculator/
10.: https://en.wikipedia.org/wiki/Common_integrals_in_quantum_field_theory